\documentclass[10pt, a4paper]{article}
\usepackage{float, graphics, subfigure, booktabs, lineno, color, authblk, multirow, makecell, fancyhdr, lastpage}
\usepackage{newtxtext,newtxmath,amsmath}
\usepackage{orcidlink}
\usepackage[super]{cite}
\usepackage[left=1.5cm, right=1.5cm, top=2.0cm, bottom=2.0cm, headsep=0.5cm]{geometry}
\usepackage[labelfont={bf,sf}, labelsep=period, justification=raggedright]{caption}

\hypersetup{colorlinks=true, allcolors=blue}
\DeclareCaptionLabelSeparator{sep}{$\ |\ $}
\makeatletter
\renewcommand\normalsize{%
\@setfontsize\normalsize\@xpt\@xiipt
\abovedisplayskip 2.5\p@ \@plus2\p@ \@minus5\p@
\abovedisplayshortskip \z@ \@plus3\p@
\belowdisplayshortskip 6\p@ \@plus3\p@ \@minus3\p@
\belowdisplayskip \abovedisplayskip
\let\@listi\@listI}
\makeatother

\title{\textbf{The local Gaussian correlation networks among return tails in the Chinese stock market}}

\author{Peng Liu \orcidlink{0000-0002-9115-2055}}
\affil{School of Information, Xi'an University of Finance and Economics, Xi'an 710100, Shaanxi, P.R. China}
\affil{\href{mailto:pengliuhep@outlook.com}{pengliuhep@outlook.com}}
\date{}

\begin{document}

\maketitle
\thispagestyle{empty}

\begin{abstract}
Financial networks based on Pearson correlations have been intensively studied.
However, previous studies may have led to misleading and catastrophic results because of several critical shortcomings of the Pearson correlation.
The local Gaussian correlation coefficient, a new measurement of statistical dependence between variables,
has unique advantages including capturing local nonlinear dependence and handling heavy-tailed distributions. 
This study constructs financial networks using the local Gaussian correlation coefficients between tail regions of stock returns in the Shanghai Stock Exchange.
The work systematically analyzes fundamental network metrics including node centrality, average shortest path length, and entropy.
Compared with the local Gaussian correlation network among positive tails and the conventional Pearson correlation network,
the properties of the local Gaussian correlation network among negative tails are more sensitive to the stock market risks.
This finding suggests researchers should prioritize the local Gaussian correlation network among negative tails.
Future work should reevaluate existing findings using the local Gaussian correlation method.

\end{abstract}

{\noindent}\rule[0pt]{18cm}{0.1em}  

\section{Introduction}
Over the past decades,
network science has been widely used to study complex systems from multiple academic fields,
and gained insightful findings that are not accessible with traditional methods.\cite{network_medicine, network_ecology, network_economy, otherfield1, otherfield2, otherfield3, network_finance, anti-network, MST}
The financial system is a typical complex system.
Therefore, various complex financial networks have been studied since 1999.\cite{network_finance, anti-network, MST}

Among various complex financial networks,
the correlation-based networks in various financial markets have been studied intensively.
\cite{network_finance, anti-network, MST, PMFG, TMFG, CBNet-2, CBNet-3, CBNet-4, CBNet-5, CBNet-6, CBNet-7, CBNet-9, CBNet-10, CBNet-11}
Such networks are crucial in understanding complex financial systems and their systemic risks.
\cite{network_finance, anti-network, MST, PMFG, TMFG, CBNet-2, CBNet-3, CBNet-4, CBNet-5, CBNet-6, CBNet-7, CBNet-9, CBNet-10, CBNet-11}
Previous works often constructed networks based on Pearson correlation coefficients $\rho$ among returns of financial asset prices.\cite{anti-network}
However, Pearson's $\rho$ has some critical shortcomings which can lead to misleading and potentially catastrophic results.\cite{locgauss-2022}

Tj$\o$stheim et. al. summarized the shortcomings of Pearson's $\rho$ as three issues.\cite{locgauss-2022}
The first one is the non-Gaussian issue.
It works well in the Gaussian case and a few non-Gaussian situations. 
However, it is unsuitable for heavy-tailed distributions,
which characterize financial asset returns.\cite{locgauss-2022, logreturn}
The second one is the robustness issue.
It is well known that Pearson's $\rho$ is sensitive to outliers.
Even one single outlier in data may cause a damaging change in Pearson's $\rho$.
The last one is the nonlinearity issue, which may be the most damaging over other shortcomings.
Consider the case $Y = X^{2}$.
Although the relationship between variables  $Y$ and $X$ is the strongest form of dependence in this case,
Pearson's $\rho$ between these two variables may be zero.
Pearson's $\rho$ fails to capture nonlinear asymmetry statistical dependence,
which widely exists among financial markets.\cite{locgauss-2022, locgauss-2013, locgauss-2014, locgauss-2020}
Therefore, Pearson's $\rho$ cannot capture the complex statistical dependence for heavy-tailed variables and nonlinear situations in financial markets.

Two stylized facts characterize financial asset returns: (1) distributions of financial asset returns exhibit heavy-tailed properties,\cite{logreturn}
and (2) correlations among asset returns demonstrate nonlinearity and asymmetry.\cite{locgauss-2022, locgauss-2013, locgauss-2014, locgauss-2020}
Given these empirical characteristics and the aforementioned limitations of Pearson's $\rho$,
its application in measuring statistical dependence between financial asset returns may yield misleading conclusions with potentially catastrophic consequences.\cite{locgauss-2022}
Consequently, previous studies on the correlation networks constructed using Pearson's $\rho$ did not accurately capture the true interconnectedness of financial systems,
and thus may have produced misleading findings.

To overcome the weaknesses of Pearson's $\rho$,
researchers have proposed alternative methods for measuring nonlinear statistical dependence,
including the copula method, the conditional correlation, the distance covariance, the HSIC (Hilbert-Schmidt Independence Criterion) measure, the mutual information, and the local Gaussian correlation, among others.
For detailed comparisons of these alternatives, this paper refers readers to Refs.~\citen{locgauss-2013, locgauss-2014, locgauss-2022}.
Among these approaches, the local Gaussian correlation proposed by Tj$\o$stheim and Hufthammer demonstrates particularly great performance.\cite{locgauss-2013, locgauss-2014, locgauss-2022}
This measure effectively captures local nonlinear asymmetry dependence and applies to heavy-tailed distributions.\cite{locgauss-2013, locgauss-2014, locgauss-2020, locgauss-2022}
As this measure has been successfully applied across diverse research domains,\cite{locgauss-2013, locgauss-2014, locgauss-2020, locgauss-2022}
this study employs this measure to analyze local nonlinear statistical dependence between stock returns in the Chinese stock market.

In finance,
the local nonlinear statistical dependence among tail regions of financial asset returns is of special interest because it plays a crucial role in understanding the market risks and optimizing investment portfolios.\cite{locgauss-2013, locgauss-2014, locgauss-2020, portfolio, portfolio2}
Therefore,
the purpose of this study is to introduce both the negative-tail and positive-tail local Gaussian correlation networks (LGCNETs)
by measuring the local Gaussian correlations among tail regions of stock returns in the Chinese stock market.
This study also aims to compare the LGCNETs with the conventional Pearson correlation networks studied previously.

The remainder of this paper is organized as follows.
Section~\ref{sec:data} gives the information on the dataset used in this analysis.
Section~\ref{sec:LGC} briefly introduces the local Gaussian correlation measure and the methodology for constructing LGCNETs among return tails.
Section~\ref{sec:metrics}  briefly introduces the network metrics analyzed in this study.
Section~\ref{sec:results} presents the empirical results and discussion.
Section~\ref{sec:conclusion} concludes this paper.

\section{Data}
\label{sec:data}
\begin{figure}[H]
	\centerline{\includegraphics[width=1.0\linewidth]{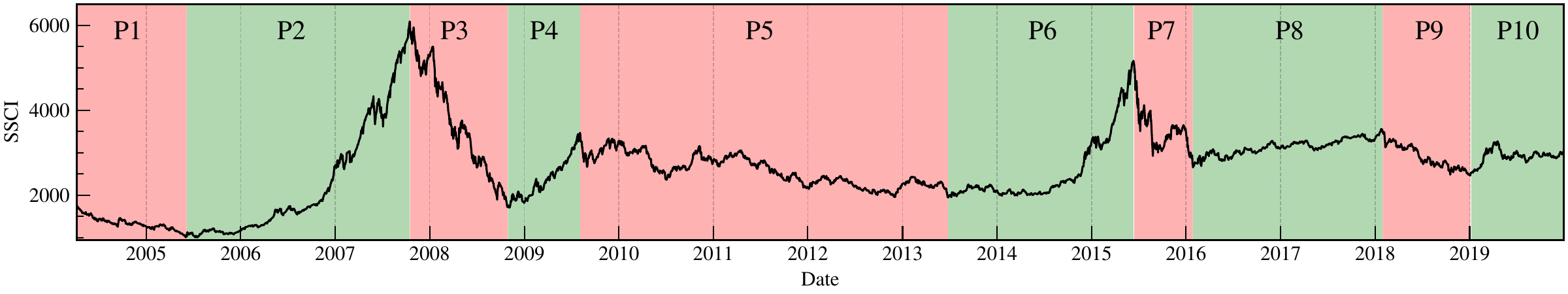}}
	\vspace*{8pt}
	\caption{SSCI performance over the period from April 7, 2004 to December 31, 2019.
	Based on SSCI fluctuations, the period is segmented into ten periods as colored regions indicate.
	Red and green regions represent bear and bull markets, respectively.}
	\label{fig1}
\end{figure}

This study analyzes the daily closing prices of all stocks traded on the main board of the Shanghai Stock Exchange (SSE) over the period from April 7, 2004 to December 31, 2019.
The dataset comprises 1542 stocks with 3639462 closing price records, sourced from Eastmoney's official website (\url{https://quote.eastmoney.com}).
Based on fluctuations in the Shanghai Securities Composite Index (SSCI), which is a benchmark reflecting the overall performance of the Chinese stock market,
this study divides the period into ten periods as shown by Fig.~\ref{fig1}.
The detailed information of each period is listed in Table~\ref{table1}.
The stocks with missing logarithmic returns on more than 30 trading days are removed from the datasets in each period.

As shown in Fig.~\ref{fig1},
the study period covers the two most significant crashes in the Chinese stock market in this century: the 2007-2008 global financial crisis and the 2015-2016 domestic stock market turbulence.
It is a common statement that such extreme market events strengthen the correlations among return tails.\cite{locgauss-2013}
These two crashes provide a unique opportunity to examine how international and domestic financial crises influence the structural dynamics of LGCNETs analyzed in this study.
\begin{table}[H]
	\centering
	\caption{Information of ten periods.
	For each period, detailed information is provided: starting date, ending date, the percentage change of SSCI, number of stocks available to construct the network, and number of trading days.}
	\label{table1}
	\begin{tabular}{cccccc}
		\toprule
		Period & Starting date & Ending date & Percentage change & Number of stocks & Number of trading days \\
		\midrule
		1      & 20040407      & 20050606    & -44\%             & 769              & 281                    \\
		2      & 20050607      & 20071016    & 514\%             & 224              & 574                    \\
		3      & 20071017      & 20081028    & -73\%             & 740              & 253                    \\
		4      & 20081029      & 20090804    & 109\%             & 812              & 189                    \\
		5      & 20090805      & 20130625    & -47\%             & 633              & 940                    \\
		6      & 20130626      & 20150612    & 180\%             & 653              & 481                    \\
		7      & 20150613      & 20160127    & -49\%             & 850              & 154                    \\
		8      & 20160128      & 20180129    & 36\%              & 796              & 490                    \\
		9      & 20180130      & 20190104    & -32\%             & 1273             & 226                    \\
		10     & 20190105      & 20191231    & 25\%              & 1446             & 241                   \\
		\bottomrule
	\end{tabular}
\end{table}

\section{Local Gaussian correlation and network construction}
\label{sec:LGC}
The local Gaussian correlation proposed by Tj$\o$stheim et. al. is a new measure of statistical dependence between variables.\cite{locgauss-2013, locgauss-2022}
This new measure effectively captures nonlinear local correlation and is suitable for heavy-tailed variables.\cite{locgauss-2013, locgauss-2022}
The key idea of this measurement is that
the Gaussian bivariate density approximates locally the empiriccal bivariate density $f$ for two varibales $\left( X_{1}, X_{2}\right)$ in neighborhood of each point $x = \left(x_{1}, x_{2}\right)$.
The Gaussian bivariate density $\psi$ is defined by Eq.~(\ref{Eq:bigaussian}).
\begin{eqnarray}
	\label{Eq:bigaussian}
		\psi \left(x \right)
		=
		\frac{1}{2\pi \sigma_{1}\left(x\right) \sigma_{2}\left(x\right) \sqrt{1 - \rho^{2}\left(x\right)}}
		 e^{
			 -\frac{1}{2\left[1 - \rho^{2}\left(x\right) \right]}
			 \left\{
			 \left[\frac{v_{1} - \mu_{1}\left(x\right)}{\sigma_{1}\left(x\right)}\right]^2 +
			 \left[\frac{v_{2} - \mu_{2}\left(x\right)}{\sigma_{2}\left(x\right)}\right]^2 -
			 2\rho\left(x\right)
			 \left[ \frac{v_{1} - \mu_{1}\left(x\right)}{\sigma_{1}\left(x\right)} \right]
			 \left[ \frac{v_{2} - \mu_{2}\left(x\right)}{\sigma_{2}\left(x\right)} \right]
			 \right\}
		 }
\end{eqnarray}
where $v_{i}$, $\mu_{i} \left( x \right)$, and $\sigma_{i} \left( x \right)$ $\left(i = 1, 2\right)$ are the running variables in the Gaussian distribution,
the local means, and the local standard deviations, respectively.
The $\rho \left( x \right)$ is the local Gaussian correlation coefficient at the point $x = \left(x_{1}, x_{2}\right)$,
which is what we need to estimate here.

The theory for estimating $\rho \left( x \right)$ has been detailed in Ref.~\citen{locgauss-2013}.
The routines for estimating $\rho \left( x \right)$ have been implemented in the R package localgauss.\cite{Rpackage}
This study uses the localgauss package and the plug-in bandwidth $b = 1.75\sigma n^{-\frac{1}{6}}$
($\sigma$ and $n$ refer to the standard deviation and the number of observations for a variable, respectively)\cite{Rpackage_lg} to estimate the local Gaussian correlation coefficient $\rho\left(x \right)$.
This paper also conducts a robustness analysis involving slight variations in bandwidth.
The robustness analysis demonstrates that the findings presented in this paper are robust to changes in bandwidth.

In finance, the correlation between financial asset return tails is of special interest because these tails contain extreme market risks.
By quantifying the local Gaussian correlations among tail regions of the stock returns in the Chinese stock market,
this study constructs LGCNETs both among negative tails and among positive tails.
For each pair of stocks in a period introduced in the previous section,
this study first measures the diagonal local Gaussian correlation,
and then calculates the mean values on the diagonal in the negative tail from quantile 5\% to 20\% and in the positive tail from quantile 80\% to 95\%, respectively.
These two mean values are used as the link weights for negative-tail LGCNET and positive-tail LGCNET, respectively.

Based on the link weights mentioned above,
this study employs the Minimum/Maximum Spanning Tree (MST),\cite{MST}
the Planar Maximally Filtered Graph (PMFG),\cite{PMFG}
and the Triangulated Maximally Filtered Graph (TMFG)\cite{TMFG} methods
to extract the most important links of LGCNETs.
For correlation-based networks,
the minimum Spanning Tree (Minimum distance between nodes) and Maximum Spanning Tree (Maximum weight of links) generate the same MST network
because a larger link weight means a smaller distance between nodes linked by that link.
These filtering methods are widely used in constructing correlation networks.\cite{filtering}
Their technical details can be found in Refs.~\citen{MST, PMFG, TMFG}.

To compare with previous studies,
the study also constructs conventional Pearson correlation networks in each period using MST, PMFG, and TMFG filtering methodologies.

\section{Network metrics}
\label{sec:metrics}
This study analyzes the evolution characteristics of the fundamental metrics of negative-tail LGCNETs, positive-tail LGCNETs,  and the Pearson correlation networks under MST, PMFG, and TMFG filtering methodologies.
These metrics include node strength centrality, node eigenvector centrality, average shortest path length, and network entropy (Shannon entropy, Rényi entropy, and Tsallis entropy).

Node centrality quantifies the importance of nodes within a network.
Among various centrality measures available,
this study selects strength centrality and eigenvector centrality for analysis based on their distinct characteristics in capturing different dimensions of node influence.\cite{anti-network, degreeandstrength, eigenvector}
In correlation-based financial networks,
both strength centrality and eigenvector centrality play crucial roles in risk management, portfolio optimization, and asset price prediction, among others.\cite{anti-network, nodecentrality1, nodecentrality2}
Their distributions are also key to understanding the error and attack tolerance of complex financial systems.\cite{scale-free-network}

For a correlation-based financial network with $N$ nodes and weight matrix $W$,
the strength centrality $s_{i}$ of node $i$ is defined as Eq.~\ref{Eq:strength},
and the eigenvector centrality $v_{i}$ is the $i$th component of the leading eigenvector $v$ of weight matrix $W$.\cite{anti-network, degreeandstrength, eigenvector}
\begin{eqnarray}
	\label{Eq:strength}
	s_{i} = \sum\limits_{j = 1}^{N} w_{ij}
\end{eqnarray}

The average shortest path length is a metric that quantifies the typical separation between two nodes in a network, and it is crucial for understanding risk propagation in complex financial systems.\cite{anti-network, CBNet-2}
For a correlation-based financial network with $N$ nodes and weight matrix $W$,
the average shortest path length $\left< L \right>$ is defined as Eq.~\ref{Eq:L}.\cite{anti-network, CBNet-2}
 \begin{eqnarray}
	\label{Eq:L}
	\left< L \right> = \frac{1}{N\left(N-1\right)}\sum\limits_{i \neq j}^{N}l_{ij}
\end{eqnarray}
where $l_{ij}$ is the shortest path length between node $i$ and node $j$.
The shortest path is the path that minimizes the total edge distance,
where the distance $d_{ij}$ between node $i$ and node $j$ is defined as $d_{ij} = \sqrt{2 \left( 1 - w_{ij} \right)}$.

Network entropy serves as a fundamental metric frequently employed in quantifying financial network resilience,
utilizing methodologies derived from statistical mechanics and ergodic theory.\cite{CBNet-2}
For a correlation-based financial network with $N$ nodes and weight matrix $W$,
the Shannon entropy HS$_{i}$, Rényi entropy HR$_{i}$, and Tsallis entropy HT$_{i}$ of node $i$ are defined as Eq.~\ref{Eq:Shannon}, Eq.~\ref{Eq:Renyi}, and Eq.~\ref{Eq:Tsallis}, respectively.\cite{CBNet-2, CBNet-11, entropy}
 \begin{eqnarray}
	\label{Eq:Shannon}
	\text{HS}_{i} = -\sum\limits_{j = 1}^{N} p_{ij} \log p_{ij}
\end{eqnarray}
 \begin{eqnarray}
	\label{Eq:Renyi}
	\text{HR}_{i} = \frac{1}{1 - \beta} \log\left(\sum\limits_{j = 1}^{N} p_{ij}^{\beta}\right)
\end{eqnarray}
 \begin{eqnarray}
	\label{Eq:Tsallis}
	\text{HT}_{i} = \frac{1}{\beta - 1} \log\left(1 - \sum\limits_{j = 1}^{N} p_{ij}^{\beta}\right)
\end{eqnarray}
 \begin{eqnarray}
	\label{Eq:pmatrix}
	p_{ij} = \frac{w_{ij}} {\sum\limits_{j = 1}^{N} w_{ij}}
\end{eqnarray}
where $p_{ij}$ is the element of transition probability matrix $P$, and is defined as Eq.~\ref{Eq:pmatrix}.
$\beta$ is an adjustable parameter.
In this study, the results are reported with $\beta$ equal to 0.1.
This study also varies $\beta$ in the interval (0, 1) to see how the results change.
Such analysis illustrates that the findings of this paper are robust to parameter $\beta$.

Based on the node entropy defined above,
the Shannon entropy HS, Rényi entropy HR, and Tsallis entropy HT of a network
are defined as Eq.~\ref{Eq:NetworkEntropy}.\cite{CBNet-2, CBNet-11, entropy}
 \begin{eqnarray}
	\label{Eq:NetworkEntropy}
	\text{HS} = \sum\limits_{i = 1}^{N} \pi_{i} \text{HS}_{i},\,\,\,\, \text{HR} =  \sum\limits_{i = 1}^{N} \pi_{i} \text{HR}_{i}, \,\,\,\, \text{HT} = \sum\limits_{i = 1}^{N} \pi_{i} \text{HT}_{i}
\end{eqnarray}
where $\pi_{i}$ is the $i$th component of the stationary distribution $\pi$ of the corresponding transition probability matrix $P$, which satisfies $\pi = \pi P$.

\section{Empirical results and discussion}
\label{sec:results}
\begin{figure}[H]
	\centerline{\includegraphics[width=1.0\linewidth]{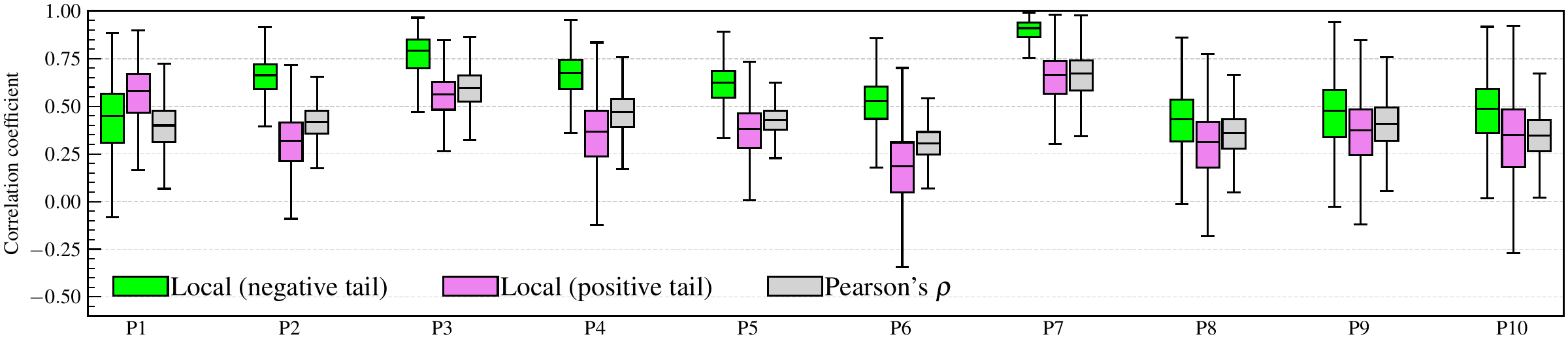}}
	\vspace*{8pt}
	\caption{Boxplots of correlation coefficients across ten periods.
	Green boxes, violet boxes, and gray boxes indicate local Gaussian correlation coefficients among negative tails, local Gaussian correlation coefficients among positive tails, and conventional Pearson correlation coefficients.}
	\label{fig2}
\end{figure}

The correlation coefficients calculated to construct networks
are directly related to the network properties.
Therefore, this paper first investigates the distributions of both the local Gaussian correlations (both among negative tails and among positive tails) and Pearson correlation coefficients.
As shown in Fig.~\ref{fig2}, which compares these distributions via boxplots, almost all coefficients are positive.
During both bull and bear periods,
the local Gaussian correlation coefficients among negative tails are statistically larger than those among positive tails.
The relatively large correlation coefficients make us pay more attention to the correlations among negative tails.
During the periods of both the 2007-2008 global financial crisis and the 2015-2016 Chinese stock market turbulence,
the market's rapid decline has significantly increased the correlation coefficients.
Compared with the global financial crisis, the domestic crash has a more significant impact on the correlations.
Notably,
Pearson's $\rho$ shows that its distributions are different from the local Gaussian correlation coefficient distributions.
This difference indicates the necessity of studying the LGCNET.

\begin{figure}[H]
	\centerline{\includegraphics[width=1.0\linewidth]{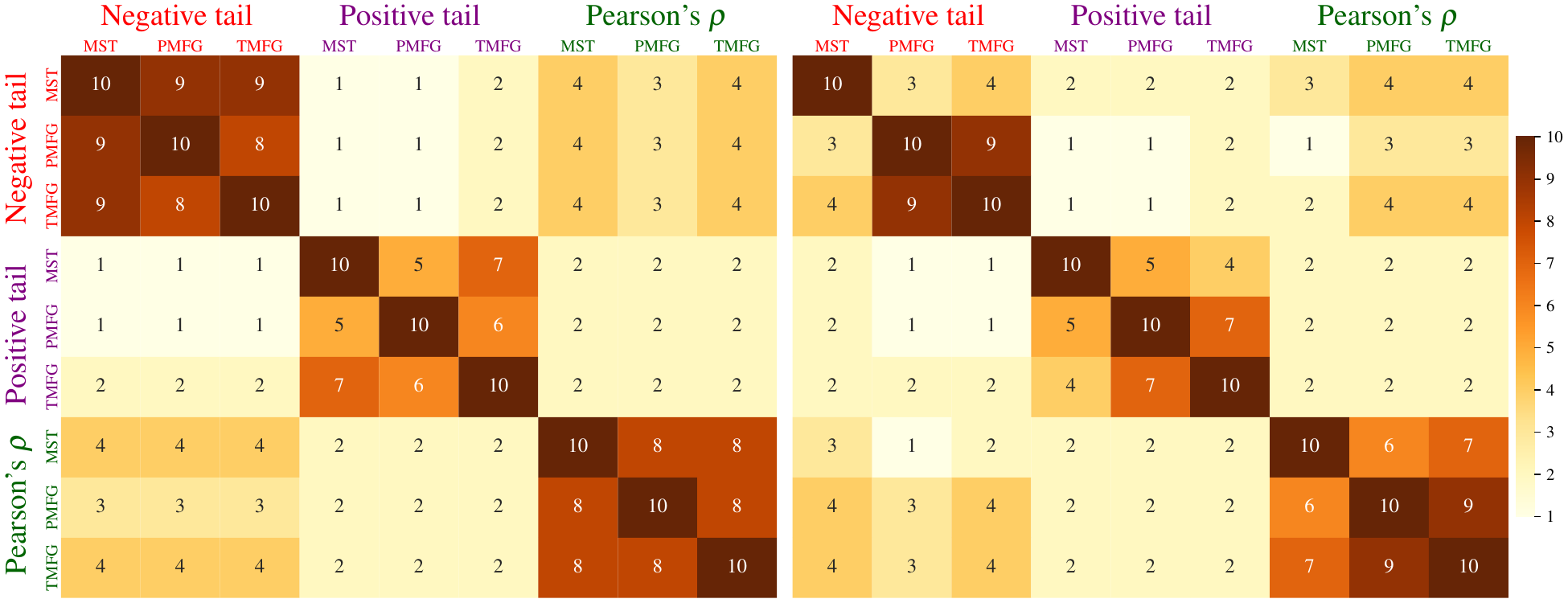}}
	\vspace*{8pt}
	\caption{Pairwise overlap of top-ten stocks in filtered correlation networks during the last period. 
	The left and right panels depict strength and eigenvector centrality rankings, respectively.
	See main text for details.}
	\label{fig3}
\end{figure}

Strength centrality and eigenvector centrality are critical metrics for assessing node importance in weighted financial correlation-based networks.\cite{anti-network, degreeandstrength, eigenvector}
This study conducts a comparative analysis of stock rankings generated by these centrality measures
across three network types: negative-tail LGCNET, positive-tail LGCNET, and conventional Pearson correlation network.
Fig.~\ref{fig3} presents the pairwise overlap of top-ten stocks among filtered variants of the three network types during the last period.
This figure illustrates two key patterns:
(1) the top-ten stocks are highly similar among filtered networks of the same type,
whereas (2) the top-ten stocks are different among the three network types.
This work also examines the comparative results in the remaining nine periods,
showing that the top-ten stocks among these three network types remain distinct in each period.
The findings of this comparative study further demonstrate the necessity of prioritizing LGCNET in analyses.

The node centrality distribution is directly related to the resilience of complex financial networks.\cite{scale-free-network}
Prior research has demonstrated that both degree and strength distributions in financial networks
follow heavy-tailed patterns—a defining topological characteristic critical for understanding the error and attack tolerance of these systems.
Therefore,
by estimating the tail shape parameters of both strength and eigenvector centrality distributions using the Generalized Pareto Distribution (GPD),\cite{GPD-1, GPD-2, GPD-3, COVID-19-1}
this study examines whether the LGCNETs analyzed here exhibit scale-free properties in terms of strength centrality and eigenvector centrality.

\begin{figure}[H]
	\centerline{\includegraphics[width=1.0\linewidth]{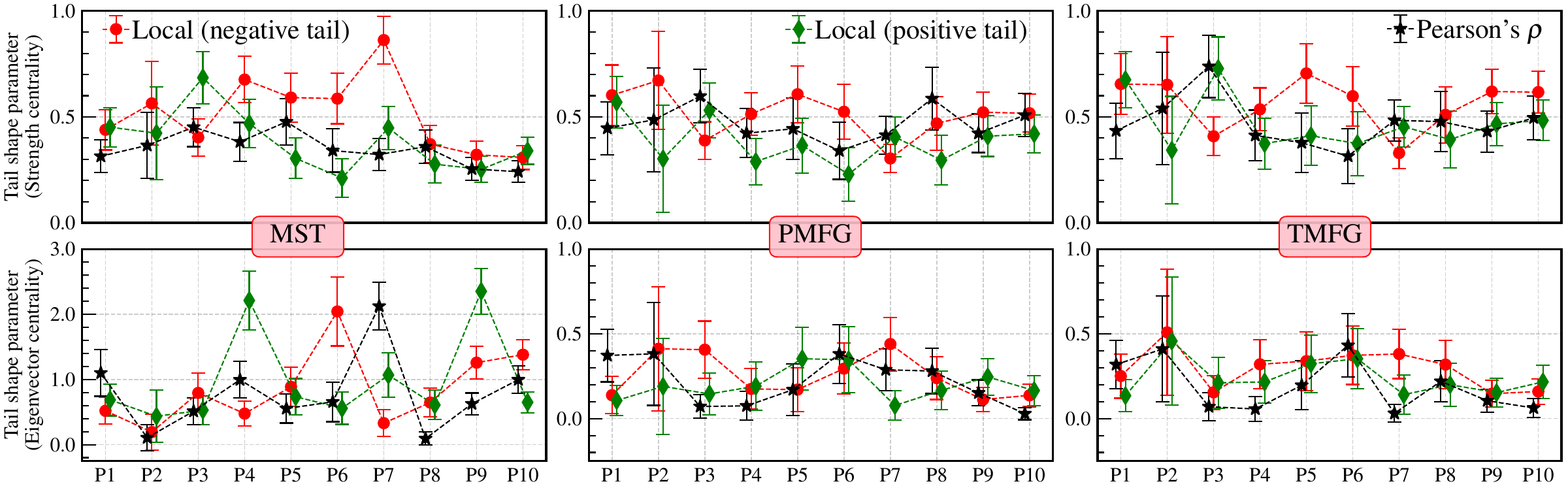}}
	\vspace*{8pt}
	\caption{Tail shape parameters of network centrality distributions in ten periods.
	The first and second rows are for strength and eigenvector centralities, respectively.
	The first, second, and third columns are for MST, PMFG, and TMFG filtering methodologies, respectively.
	Red, green, and black data markers are for negative-tail LGCNETs, positive-tail LGCNETs, and Pearson correlation networks, respectively.
	Vertical error bars are 95\% confidence intervals.
	These panels share common legends.}
	\label{fig4}
\end{figure}

The GPD estimations are shown in Fig.~\ref{fig4}.
From this figure, we see that all shape parameters for LGCNETs and Pearson correlation networks are positive.
A positive tail shape parameter indicates a heavy-tailed distribution.
This means that all networks analyzed here are scale-free in terms of both strength centrality and eigenvector centrality.
The figure further demonstrates statistically indistinguishable tail shape parameters across the three network types
 under both PMFG and TMFG filtering methodologies.
For the MST filtering methodology, however,
the tail shape parameters of the three network types are statistically different.

\begin{figure}[H]
	\centerline{\includegraphics[width=1.0\linewidth]{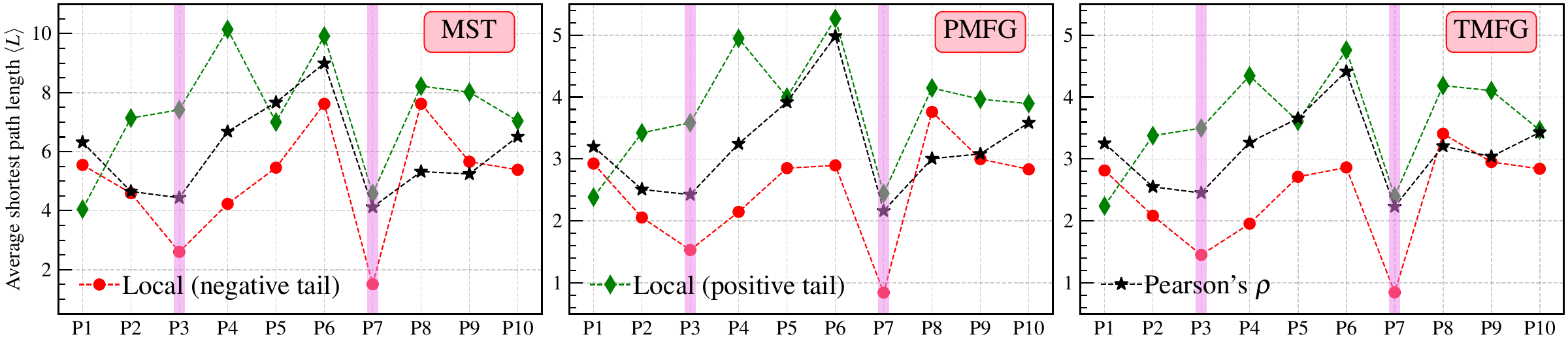}}
	\vspace*{8pt}
	\caption{Network's average shortest path length across ten periods.
	The left, middle, and right panels are for MST, PMFG, and TMFG filtering methodologies, respectively.
	Red, green, and black data markers are for negative-tail LGCNETs, positive-tail LGCNETs, and Pearson correlation networks, respectively.
	Violet vertical bands indicate the periods of the 2007-2008 global financial crisis and the 2015-2016 Chinese stock market turbulence.
	These panels share common legends.}
	\label{fig5}
\end{figure}

Average shortest path length $\left< L \right>$ quantifies the typical distance between two nodes in a complex network.\cite{anti-network, CBNet-2}
This observation can reflect the ability of risk propagation in a financial network.\cite{anti-network, CBNet-2}
Fig.~\ref{fig5} presents this measurement for the LGCNETs and Pearson correlation networks across ten periods.
It shows that the $\left< L \right>$ values of each network type under three filtering methodologies exhibit similar temporal changing trends. 
Compared with the positive-tail LGCNETs and the Pearson correlation networks, the $\left< L \right>$ values for the negative-tail LGCNETs are smaller in almost all periods.
The extreme decline events during short periods,
i.e., the 2007--2008 global financial crisis and the 2015--2016 Chinese stock market turbulence,
have significantly decreased the $\left< L \right>$.
Notably, the domestic market crash made the $\left< L \right>$ the smallest.
Such findings are in agreement with the results presented in Fig.~\ref{fig1}.
For all three filtering methodologies, the $\left< L \right>$ values of Pearson correlation networks and the negative-tail LGCNETs have a similar changing trend.
However, their values are different.
If we use Pearson correlation networks to estimate the ability of risk propagation in financial complex systems,
the ability will be underestimated.
The findings stated above illustrate that
the negative-tail LGCNETs have a stronger ability in risk propagation and we should pay more attention to them.

\begin{figure}[H]
	\centerline{\includegraphics[width=1.0\linewidth]{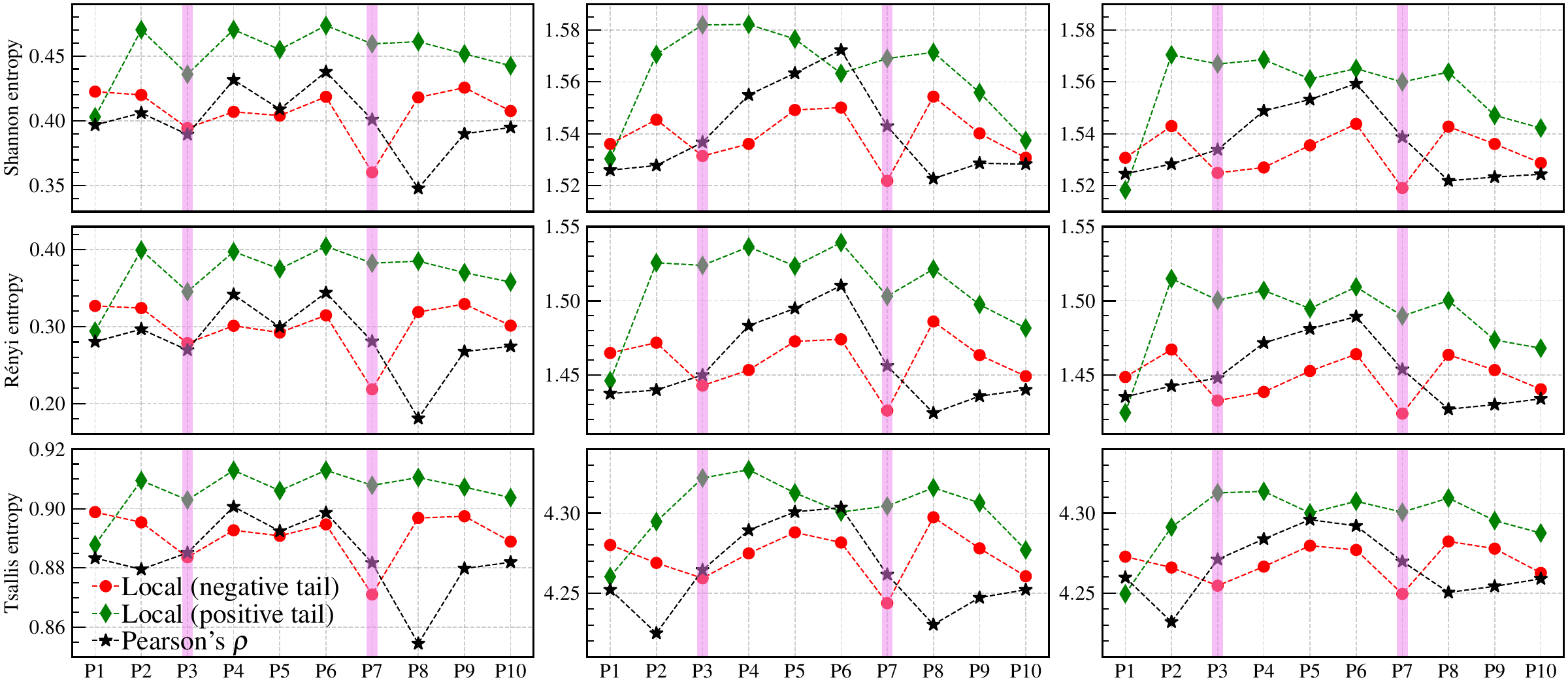}}
	\vspace*{8pt}
	\caption{Network's entropy across ten periods.
	The first, second, and third rows show Shannon entropy, Rényi entropy, and Tsallis entropy, respectively.
	The first, second, and third columns are for MST, PMFG, and TMFG filtering methodologies, respectively.
	Red, green, and black data markers are for negative-tail LGCNETs, positive-tail LGCNETs, and Pearson correlation networks, respectively.
	Violet vertical bands indicate the periods of the 2007-2008 global financial crisis and the 2015-2016 Chinese stock market turbulence.
	These panels share common legends.}
	\label{fig6}
\end{figure}

Network entropy is related to network resilience.\cite{CBNet-2, CBNet-11, entropy}
Higher network entropy generally correlates with enhanced network resilience.
This study systematically analyzes the  Shannon entropy, Rényi entropy, and Tsallis entropy of negative-tail LGCNETs, positive-tail LGCNETs, and conventional Pearson correlation networks.
Fig.~\ref{fig6} shows the three entropy measurements of these three network types under MST, PMFG, and TMFG filtering methodologies across ten periods.
Under each filtering methodology,
we see from this figure that the Shannon entropy, Rényi entropy, and Tsallis entropy of negative-tail LGCNETs are all smaller than the corresponding entropies of positive-tail LGCNETs in almost all periods.
This demonstrates that the resilience of negative-tail LGCNETs is worse than that of positive-tail LGCNETs.
Therefore, we should pay more attention to the negative-tail LGCNETs.
The 2007--2008 global financial crisis had no significant impact on the entropy of both negative-tail and positive-tail LGCNETs.
However, the 2015--2016 Chinese stock market turbulence had a significant impact on the entropy of negative-tail LGCNETs.
The domestic market crash significantly decreased the entropy of negative-tail LGCNETs,
and thus dramatically changed the resilience of negative-tail LGCNETs.
Notably,
The smallest entropy for the negative-tail LGCNETs appears in period seven and for the Pearson correlation networks appears in period eight.
This demonstrates that Pearson correlation networks cannot capture the risks of stock market crashes.

\section{Conclusions}
\label{sec:conclusion}
The correlation-based financial networks have been intensively studied.
Previous studies used to construct networks based on the Pearson correlation coefficient.
These studies may have led to misleading and catastrophic results because of several critical shortcomings of the Pearson correlation coefficient.
Compared with the Pearson correlation coefficient,
the Local Gaussian correlation coefficient, a new measurement of statistical dependence between variables,
has some new features that can capture local nonlinear correlation and are suitable for heavy-tailed variables.

In finance,
it is of particular interest to look at the local correlations among return tails of financial assets.
Therefore,
this study constructs both negative-tail and positive-tail LGCNETs using the local Gaussian correlation coefficients among tail regions of stock returns in the SSE.
For ease of comparison,
this paper also constructs the networks based on Pearson correlation coefficients.
The time of stock data analyzed in this paper covers the 2007-2008 global financial crisis and the 2015-2016 Chinese stock market turbulence,
which allows us to investigate the effects of the international and domestic market crashes.

This paper compares negative-tail LGCNET, positive-tail LGCNET, and Pearson correlation networks regarding
the tail shape parameters of both strength centrality and eigenvector centrality distributions,
the average shortest path length,
the Shannon entropy, the Rényi entropy, and the Tsallis entropy.
Compared with the positive-tail LGCNET and Pearson correlation network,
these measurements suggest that the negative-tail LGCNET better reflects the stock market risks.
Compared with the 2007-2008 global financial crisis,
the 2015-2016 Chinese stock market turbulence had a stronger impact on the negative-tail LGCNET.
Therefore, we must pay more attention to the negative-tail LGCNET and the domestic crash.

The correlation-based network approach has been applied in various financial markets and has obtained some insightful results.
Future work should reevaluate existing findings using the local Gaussian correlation method.

\section*{Acknowledgements}
This work was supported by the Humanities and Social Sciences Youth Foundation of the Ministry of Education of China [grant number 22YJCZH107];
the Shaanxi Science and Technology Department, P.R. China [grant number 2023-JC-QN-0093].
The author would like to thank the reviewers for their valuable comments which make this paper better.

\section*{Declaration of competing interest}
The author declares no competing interests.

\section*{Data availability}
The data are publicly available on the Eastmoney website (\url{https://quote.eastmoney.com}) or from the corresponding author upon reasonable request.

\bibliography{refs}

\end{document}